\documentclass[aps,prl,reprint,groupedaddress,showpacs,amsmath,amssymb,floatfix]{revtex4-1}
\usepackage{graphicx}
\usepackage{bm}
\usepackage{color}
\usepackage[normalem]{ulem}
\usepackage{amsmath}
\usepackage{braket}
\usepackage{xspace}

\usepackage[dvipsnames]{xcolor}

\usepackage[hidelinks]{hyperref}
\usepackage{cleveref}

\newcommand{\wse}{WSe\textsubscript{2}\xspace}
\newcommand{\degree}{$^\circ$\xspace}
\newcommand{\moire}{moir\'e\xspace}

\newcommand{\rucl}{$\alpha$-RuCl$_3$\xspace}
\newcommand{\twse}{tWSe$_{2}$\xspace}

\newcommand{\bvec}[1]{\boldsymbol{#1}}
\newcommand{\vdagger}{{\vphantom{\dagger}}}

\begin{document}

\title{Angle evolution of the superconducting phase diagram in twisted bilayer WSe\textsubscript{2}}

\author{Yinjie Guo$^{1}$}
\author{John Cenker$^{1}$}
\author{Ammon Fischer$^{2}$}
\author{Daniel Mu\~noz-Segovia$^{1}$}
\author{Jordan Pack$^{1}$}
\author{Luke Holtzman$^{3}$}
\author{Lennart Klebl$^{4}$}
\author{Kenji Watanabe$^{5}$}
\author{Takashi Taniguchi$^{6}$}
\author{Katayun Barmak$^{3}$}
\author{James Hone$^{7}$}
\author{Angel Rubio$^{2,8}$}
\author{Dante M. Kennes$^{2,9}$}
\author{Andrew J. Millis$^{1,8}$}
\author{Abhay Pasupathy$^{1,10}$}
\author{Cory R. Dean$^{1}$}
\email{Email: cd2478@columbia.edu}

\affiliation{$^{1}$Department of Physics, Columbia University, New York, NY 10027, USA}
\affiliation{$^{2}$Max Planck Institute for the Structure and Dynamics of Matter, Hamburg, Germany} 
\affiliation{$^{3}$Department of Applied Physics and Applied Mathematics, Columbia University, New York, New York 10027, United States}
\affiliation{$^{4}$Institut für Theoretische Physik und Astrophysik and Würzburg-Dresden Cluster of Excellence ct.qmat, Universität Würzburg, 97074 Würzburg, Germany}
\affiliation{$^{5}$Research Center for Electronic and Optical Materials, National Institute for Materials Science, 1-1 Namiki, Tsukuba 305-0044, Japan}
\affiliation{$^{6}$Research Center for Materials Nanoarchitectonics, National Institute for Materials Science,  1-1 Namiki, Tsukuba 305-0044, Japan}
\affiliation{$^{7}$Department of Mechanical Engineering, Columbia University, New York, NY 10027, USA}
\affiliation{$^{8}$Center for Computational Quantum Physics, Flatiron Institute; New York, USA}
\affiliation{$^{9}$Institute for Theory of Statistical Physics, RWTH Aachen University, and JARA Fundamentals of Future Information Technology, 52062 Aachen, Germany}
\affiliation{$^{10}$Condensed Matter Physics and Materials Science Division, Brookhaven National Laboratory; Upton, USA}
\affiliation{$^{*}$\normalfont Corresponding author. Email: cd2478@columbia.edu}

\date{\today}

\maketitle

\textbf{Recent observations of superconductivity in twisted bilayer \wse\cite{xiaSuperconductivityTwistedBilayer2025,guoSuperconductivity50degTwisted2025} have extended the family of \moire superconductors beyond twisted graphene~\cite{caoCorrelatedInsulatorBehaviour2018,yankowitzTuningSuperconductivityTwisted2019,luSuperconductorsOrbitalMagnets2019,parkTunableStronglyCoupled2021,haoElectricFieldTunable2021,zhouSuperconductivityRhombohedralTrilayer2021,zhouIsospinMagnetismSpinpolarized2022, zhangEnhancedSuperconductivitySpin2023,liTunableSuperconductivityElectron2024a,holleisNematicityOrbitalDepairing2025,choiSuperconductivityQuantizedAnomalous2025,pattersonSuperconductivitySpinCanting2025,hanSignaturesChiralSuperconductivity2025}. In \wse two different twist angles were studied, 3.65\degree\cite{xiaSuperconductivityTwistedBilayer2025} and 5.0\degree\cite{guoSuperconductivity50degTwisted2025}, and two seemingly distinct superconducting phase diagrams were reported, raising the question of whether the superconducting phases in the two devices share a similar origin. Here we address the question by experimentally mapping the evolution of the phase diagram across devices with twist angles spanning the range defined by the initial reports, and comparing the results to twist angle-dependent theory. We find that the superconducting state evolves smoothly with twist angle and at all twist angles is proximal to a Fermi surface reconstruction with, presumably, antiferromagnetic ordering, but is neither necessarily tied to the Van Hove singularity,  nor to the half band insulator. Our results connect the previously distinct phase diagrams at 3.65\degree and 5\degree, and offer new insight into the origin of the superconductivity in this system and its evolution as the correlation strength increases. More broadly, the smooth phase diagram evolution, repeatability between different devices, and dynamic gate tunability within each device, establish twisted transition metal dichalcogenides as a unique platform for the study of correlated phases as the ratio of interaction strength to bandwidth is varied.}

\begin{figure*}
\includegraphics[width=0.9\linewidth]{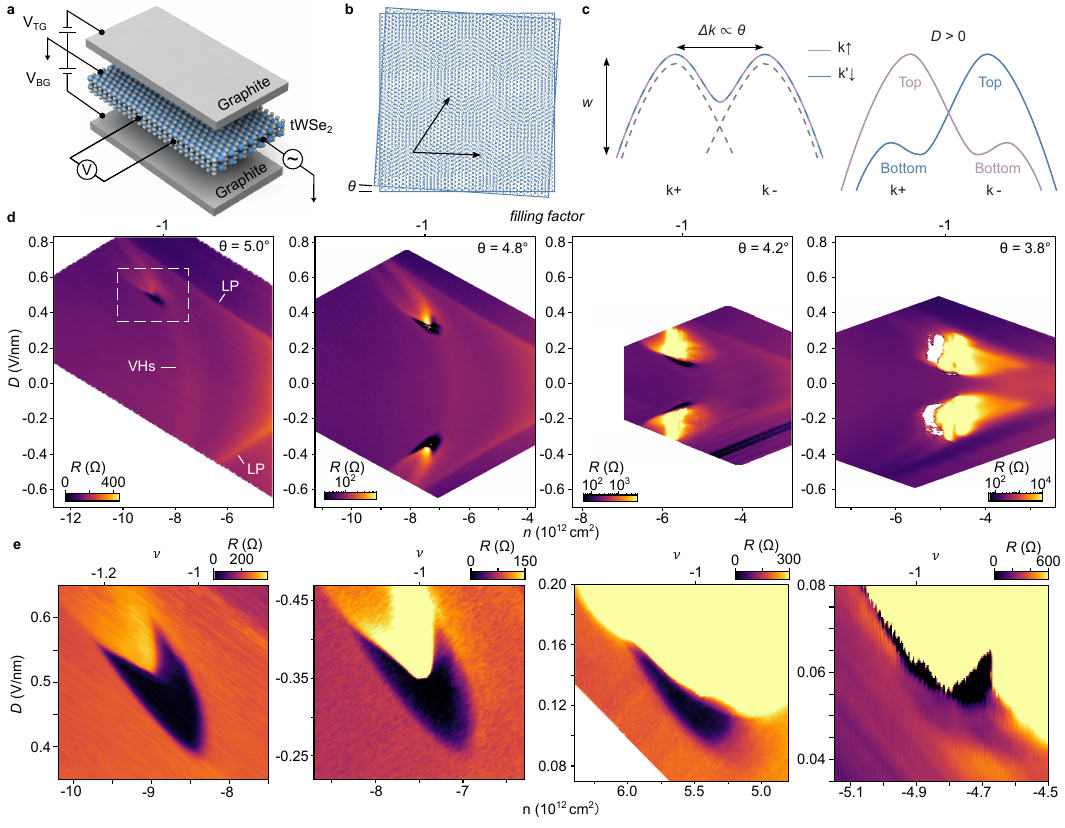}
\caption{{\bf{Twist angle evolution of the phase diagram}} 
\textbf{a} Illustration of the \twse device structure with dual gates. \textbf{b} Moir\'e superlattice formed by two layers of \wse monolayers rotated relatively to each other. \textbf{c} Illustration of band structures at zero and finite displacement field in \twse. 
\textbf{d} Density and displacement field dependence of resistance measured in \twse with twist angle at 5.0\degree, 4.8\degree, 4.2\degree, 3.8\degree. 
\textbf{e} High resolution maps of regions around the superconducting pocket. All measurements were taken at dilution fridge base temperature.}
\label{fig:fig1}
\end{figure*}

\vspace{\baselineskip}
\noindent\textbf{Introduction}

Flat band \moire systems based on transition metal dichalcogenides (TMD) exhibit a rich variety of strongly correlated states including Mott insulators~\cite{wangCorrelatedElectronicPhases2020a, tangSimulationHubbardModel2020}, generalized Wigner crystals~\cite{reganMottGeneralizedWigner2020a,liImagingTwodimensionalGeneralized2021,xuCorrelatedInsulatingStates2020}, magnetic order~\cite{andersonProgrammingCorrelatedMagnetic2023b}, and integer~\cite{liQuantumAnomalousHall2021a,fouttyMappingTwisttunedMultiband2024b} and fractional chern insulators
~\cite{caiSignaturesFractionalQuantum2023a,zengThermodynamicEvidenceFractional2023,parkObservationFractionallyQuantized2023a,xuObservationIntegerFractional2023c}. Recently, superconductivity was observed in two separate studies of twisted bilayer \wse (\twse), one at twist angle 5\degree~\cite{guoSuperconductivity50degTwisted2025},  and the other at 3.65\degree~\cite{xiaSuperconductivityTwistedBilayer2025}.

The ground state phase diagrams observed at the two twist angles differ in significant ways. At 3.65\degree~\cite{xiaSuperconductivityTwistedBilayer2025}, superconductivity appears at half filling of the twist-induced flat band  ($\nu=1$ hole per \moire unit cell), in the low displacement field regime. With increasing displacement field the superconductivity gives way to a correlated insulator, interpreted as a Mott state. 
It was argued that the superconductivity and associated phase diagram showed strong similarities to the high-T$_c$ cuprate superconductors \cite{xiaSuperconductivityTwistedBilayer2025,xiaSimulatingHightemperatureSuperconductivity2025}. In the 5.0\degree sample~\cite{guoSuperconductivity50degTwisted2025}, superconductivity emerged at large displacement fields and at  densities larger than $\nu = 1$. The superconductivity occurred near the Van Hove singularity (VHs), adjacent to a metallic antiferromagnetic (AFM) state. In this device no insulating gap was observed at half filling, and the superconductivity was interpreted to be in a moderate correlation strength regime, favouring a BCS type pairing.  Association with the magnetic transition suggested a spin fluctuation origin of the superconducting phase, similar to the situation reported in Bernal bilayer and rhombohedral multilayer graphene systems~\cite{zhouSuperconductivityRhombohedralTrilayer2021,zhangEnhancedSuperconductivitySpin2023,arpIntervalleyCoherenceIntrinsic2024,pattersonSuperconductivitySpinCanting2025,yangImpactSpinOrbitCoupling2025,holleisNematicityOrbitalDepairing2025,zhangTwistprogrammableSuperconductivitySpin2025,liTunableSuperconductivityElectron2024a,hanSignaturesChiralSuperconductivity2025,choiSuperconductivityQuantizedAnomalous2025}. 

Theoretical studies have proposed that superconductivity in \twse could have exotic character including chiral, mixed-parity, and topological order~\cite{fischerTheoryIntervalleycoherentAFM2024,xieSuperconductivityTwisted$mathrmWSe_2$2025,qinTopologicalChiralSuperconductivity2025,kleblCompetitionDensityWaves2023}, with pairing potentially arising from spin fluctuations~\cite{kleblCompetitionDensityWaves2023,fischerTheoryIntervalleycoherentAFM2024,chubukovQuantumCriticalitySuperconductivity2025,qinTopologicalChiralSuperconductivity2025,yangDisplacementFieldDrivenTransitionSuperconductivity2025}, the intersection of spin fluctuations and topology~\cite{christosApproximateSymmetriesInsulators2025,xieSuperconductivityTwisted$mathrmWSe_2$2025}, or other electron-boson  interactions~\cite{zhuSuperconductivityTwistedTransition2025}.  The impact of the twist angle on superconductivity in \twse remains unknown and it is not clear if the two experimentally observed superconductors share the same origin or represent distinct states unconnected from one another.

Here, we systemically study the evolution of the magnetic and superconducting states, as a function of twist angle, carrier density and displacement field, by mapping the transport response of tWSe\textsubscript{2} devices  with twist angles ranging from $5$\degree to $3.8$\degree. We find that the region of Fermi surface reconstruction previously observed at 5\degree, and associated with AFM ordering, continuously shifts to lower displacement field and lower filling with decreasing angle. When this pocket intersects half-filling, the Fermi surface becomes fully gapped, evidenced by activated transport. At all angles, the superconducting pocket appears at the AFM phase boundary, and correspondingly shifts to lower band filling as the twist angle decreases. The superconducting transition temperature smoothly decreases with decreasing twist angle, and at the lowest twist angles measured appears to be fully disconnected from the VHs. The observed systematic evolution of properties with twist angle combined with a detailed correspondence to theory places the $5-3.8$\degree twist angle range of tWSe\textsubscript{2} in the crossover between weak and intermediate/strong coupling.  

\begin{figure*}
\includegraphics[width=0.9\linewidth]{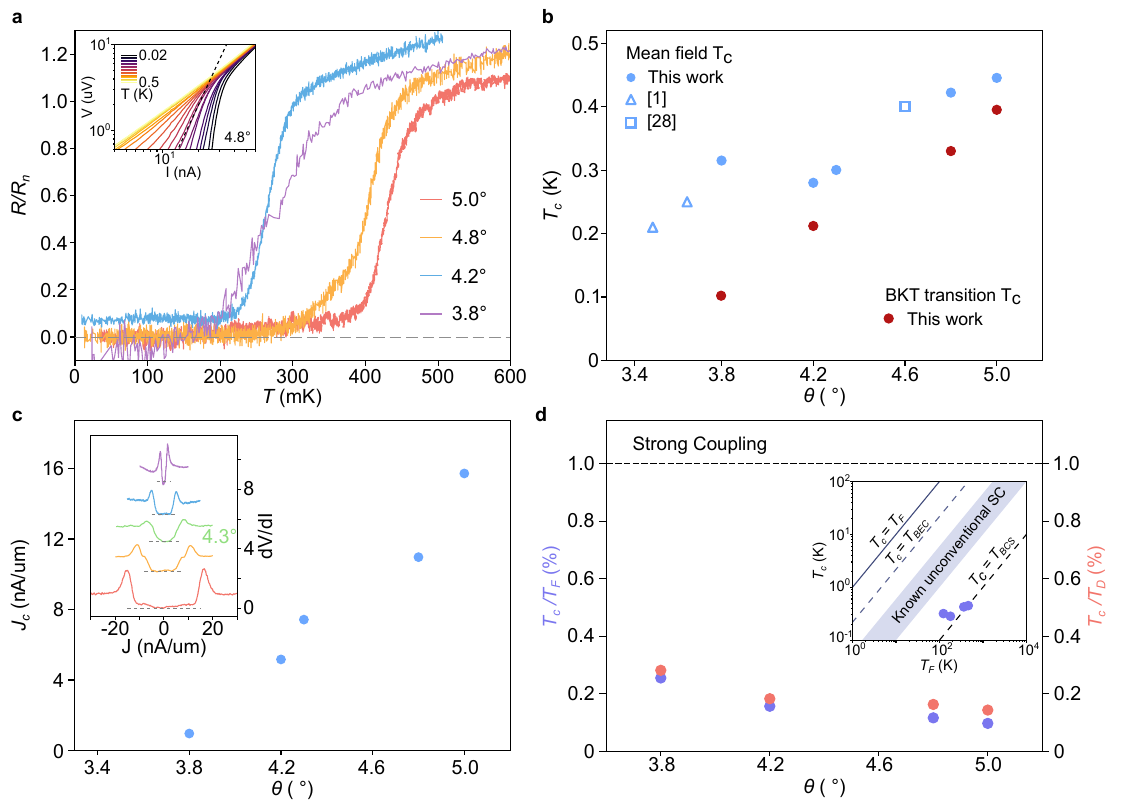}
\caption{{\bf{Superconducting properties}} \textbf{a}, Resistance versus temperature at different twist angles, measured at the density and displacement field corresponding to the highest critical temperature. 5.0\degree: -8.93$\times$ 10$^{12}$ cm$^{-2}$, 0.50~V/nm; 4.8\degree: -7.60$\times$ 10$^{12}$ cm$^{-2}$, 0.37~V/nm; 4.2\degree: -5.53$\times$ 10$^{12}$ cm$^{-2}$, -0.12~V/nm; 3.8\degree: -4.75$\times$ 10$^{12}$ cm$^{-2}$, 0.06~V/nm. Inset shows $I-V$ measurement as a function of temperature in 4.8\degree sample. The dashed line marks $V \propto I^3$. \textbf{b}, Highest critical temperature $T_c$ (defined as the $80$\% point on the resistance curve) of the superconductors observed in \twse with different twist angles and BKT temperature $T_{BKT}$ (defined as the temperature at which $V\propto I^3$) measured at the density of highest $T_c$.  Solid symbols are extracted from panel \textbf{a}. Open symbols represents data from Ref.~\cite{xiaSuperconductivityTwistedBilayer2025,xiaSimulatingHightemperatureSuperconductivity2025}. \textbf{c}, Critical current density $J$ extracted from the coherence peaks of the $dV/dI$ vs $I_{dc}$ measurement. Inset shows differential resistance $dV/dI$ as a function of d.c. current bias $I_{d.c.}$ measured in different devices. \textbf{d}, $T_c/T_F$ and $T_c/T_D$, where $T_F$ is the Fermi temperature (defined as the energy difference of the chemical potential from the valence band maximum) and $T_D$ is the temperature associated to the Drude weight, both obtained from the calculated band structure. The dashed marks where the ratio is 1\%, above which the superconductors have strong coupling. Inset shows where the superconductors fall on the Uemura plot.} 
\end{figure*} 

\vspace{\baselineskip}
\noindent\textbf{Twist angle evolution}

We measured twisted \wse in a dual gate geometry (illustrated in Fig.~1a),  and with independently gated contacts, using a fabrication process described previously~\cite{guoSuperconductivity50degTwisted2025} (See also Methods).  To realize the twisted bilayer, two \wse monolayers are laminated at a relative twist angle $\theta$, forming a \moire superlattice with  wavelength $a_M = a/\sqrt{2 (1-\cos\theta)}$ where $a = 0.328$~nm  is the lattice constant of \wse (Fig.~1b).   The valence bands from the two \wse layers hybridize as illustrated in Fig.~1c to yield a miniband with a width that decreases with twist angle, such that  smaller twist angle devices have narrower bands and stronger correlations. At zero displacement field the resulting low energy mini-bands retain a two-fold degeneracy, characterized by a coupled spin-valley pseudospin, with each band equally distributed across the two layers. Applying a displacement field lifts the degeneracy, drives a layer imbalance and further distorts the band structure (Fig.~1c)~\cite{wuTopologicalInsulatorsTwisted2019,wangCorrelatedElectronicPhases2020a,crepelBridgingSmallLarge2024a}. Thus, the electronic phase in dual-gated \twse can be experimentally tuned by three parameters: twist angle,  displacement field and carrier density.

Fig.~1d shows the longitudinal resistance measured over a wide range of  density, $n$, and displacement field, $D$, in four \twse samples with twist angles  5.0\degree , 4.8\degree , 4.2\degree and 3.8\degree (two additional devices, corresponding to twist angle 4.3\degree and 4.1\degree are shown in Figs.~SI~2 and~3). In the 5\degree panel, we label the same resistive features reported previously for this device including the layer polarization boundary, the VHs, and the region of Fermi surface reconstruction (dashed box), associated with the emergence of AFM order and superconductivity~\cite{guoSuperconductivity50degTwisted2025}. 

As the twist angle is decreased, the Fermi surface reconstruction continuously shifts to smaller displacement field, and lower filling fraction.  At the same time, the excess resistance within this pocket grows larger (note the difference in color scales). At 4.2\degree and 3.8\degree, the Fermi surface reconstruction overlaps with $\nu=1$ filling, and in this range a fully insulating state ($d\rho/dT<0$ down to lowest temperature) is observed, centered at $\nu=1$ (see $\rho$ vs $T$ in Fig.~3e). 
For the 5.0\degree and 4.8\degree devices, the magnetotransport within the Fermi surface reconstruction region (see~\cite{guoSuperconductivity50degTwisted2025} and Fig.~SI~4) is consistent with antiferromagnetic ordering with an incompletely gapped Fermi surface~\cite{ghiottoStonerInstabilitiesIsing2024,guoSuperconductivity50degTwisted2025,xiaSimulatingHightemperatureSuperconductivity2025,zangHartreeFockStudyMoire2021}. At 4.2\degree and 3.8\degree, we again see evidence for AFM ordering (see Fig.~SI~5), but in this case with a fully gapped Fermi surface. 

In each device, superconductivity is observed in a localized region that resides on the low $D$ boundary of the AFM phase.  Fig.~1e shows high resolution maps of the superconducting pockets measured from the same devices in Fig.~1d.  As the twist angle decreases, the superconducting pocket shifts along with the AFM state to lower filling fraction, reduces to a smaller sized pocket, and exhibits a decreasing critical temperature (see Fig.~2 and associated text).  
The superconducting pocket appears whether or not the Fermi surface is gapped within the AFM region, and at smaller twist angles is not associated with proximity to the VHs (see Fig.~SI~6).  

\begin{figure*}
\includegraphics[width=0.9\linewidth]{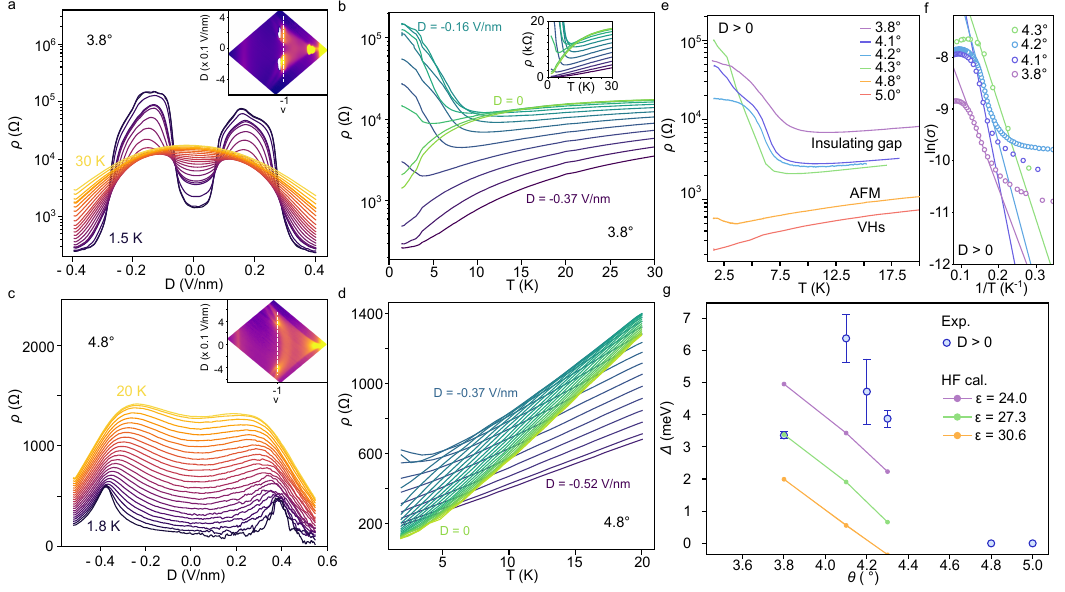}
\caption{{\bf{Insulating gaps at half filling}} \textbf{a} Temperature dependence of resistivity at half filling density varying displacement field on 3.8\degree \twse. Inset shows the line cut on the 1.5~K phase diagram where the measurement is taken at. \textbf{b} The resistivity plotted versus temperature showing an insulating behavior at intermediate displacement field range. Inset shows the same plot on linear scale. \textbf{c,d} The same measurements on 4.8\degree \twse. \textbf{e}, Resistivity $\rho$ as a function of temperature at half filling. \textbf{f}, Arrhenius plot at the half filling insulating state in low twist angle devices. Logarithm of conductivity $\sigma$ versus 1/$T$ at the displacement field where the gap size is maximum. The linear region is fit to $\sigma \sim e^{{-\Delta}/{2k_BT}}$. \textbf{g}, Gap size $\Delta$ of the insulating state at half filling extracted from \textbf{e}. The value is set to be zero for 4.8\degree and 5.0\degree devices since no fully opened gaps are observed. The solid dots and curves show the calculated gap size from Hartree Fock using different dielectric constant value $\varepsilon$.} 
\label{fig:fig3}
\end{figure*} 

\begin{figure*}
\includegraphics[width=0.9\linewidth]{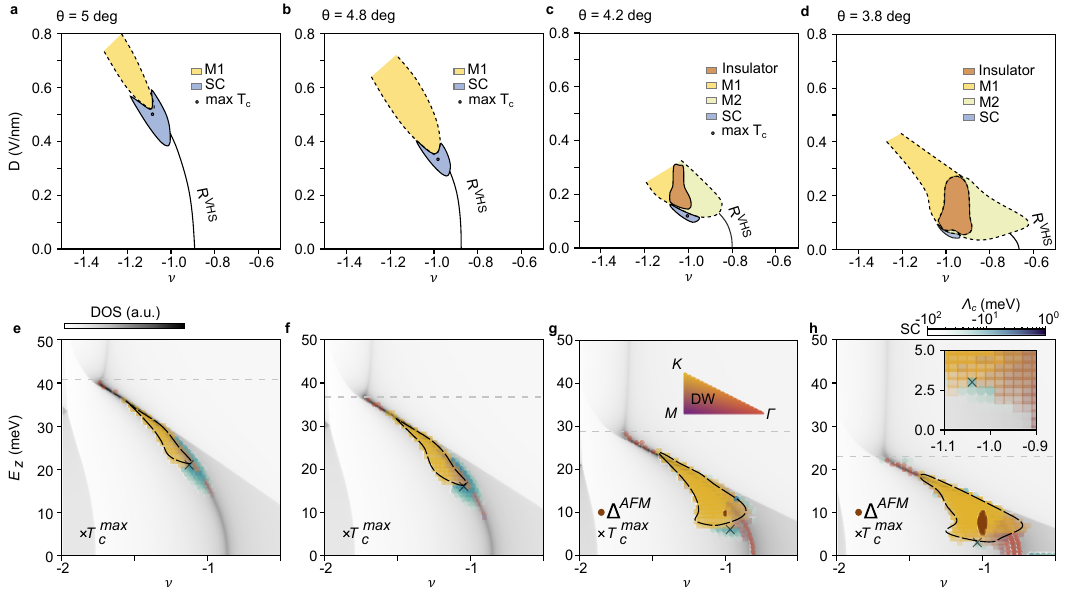}
\caption{{\bf{The superconductor and the IVC-AFM ordering}.} 
\textbf{a-d} Illustration of the phase diagram at positive displacement field, representing the relationship between the superconductors and the neighboring resistive ordering states. \textbf{e-h} Functional renormalization
group (FRG) calculations of the phase diagram as a function of displacement field and filling factor simulated for the same twist angles compared to the experimental data. The critical RG scale $\Lambda_{\mathrm{c}}$ indicates the onset temperature of SC (blue) and magnetic ordering (orange-red) in the phase diagram and is encoded in the opacity of the respective colormaps. The maximal critical temperature $T_\mathrm{c}^{\text{max}}$ of the SC pocket is marked by the cross. Different magnetic orderings (orange-red) are distinguished by the leading momentum transfer of the magnetic susceptibility $\chi^{\text{DW}}(\bvec q)$ ($K$: IVC-AFM order, $\Gamma$: valley-polarized order and $M$: striped-IVC). The black dashed curve defines the boundary of the IVC-AFM order. Solid brown points indicate the size of the correlated gap at half-filling $\nu=-1$ as predicted by Hartree-Fock. The DOS is plotted in the background to indicate the position of the VHs. FRG calculations are done below the grey dashed lines, indicating the experimentally relevant parameter regime.}
\label{fig:fig4}
\end{figure*} 

\vspace{\baselineskip}
\noindent\textbf{Superconductivity versus twist angle}

In Fig.~2 we examine in greater detail the twist angle evolution of the superconducting response.  Fig.~2a shows characteristic resistance versus temperature measurements for the same four devices shown in Fig.~1.  The curves were measured at an $n$ and $D$ value corresponding to the position of maximum superconducting $T_{c}$ for that device (see Method).

Fig.~2b plots the maximum superconducting $T_c$ (defined as the temperature at which the resistance dropped to 80\% of the normal-state value) versus twist angle.  Solid symbols are obtained from the data sets in Fig.~2a, and open symbols correspond to measurements reported in Refs~\cite{xiaSuperconductivityTwistedBilayer2025} and~\cite{xiaSimulatingHightemperatureSuperconductivity2025}. $T_{c}$ shows a remarkably smooth evolution, decreasing continuously with decreasing twist angle, and dropping by approximately a factor 2 between 5\degree and 3.5\degree.  Red circles show the BKT transition temperature, $T_{BKT}$, defined as the temperature at which $V\propto I^3$ ~\cite{tinkhamIntroductionSuperconductivity2004}. $T_{BKT}$ decreases monotonically with decreasing twist angle,  following closely the mean field $T_c$ trend. We note that the transition width for the 3.8\degree device is significantly broader than the others (Fig.~2a), and there is an increased difference between the $T_c$ and $T_{BKT}$ values (Fig.~2b). We conjecture this is due to the presence of twist angle disorder in the 3.8\degree device.   

Fig.~2c shows the critical current density, $J_{c}$, measured for each superconductor. Inset shows $dV/dI$ vs $I_{dc}$, acquired at temperature $T=20$~mK (as measured by the fridge thermometer). The differential resistance is normalized to the normal state resistance and each curve is offset for clarity (horizontal dashed line identifies the zero value for each curve).  We estimate $J_{c}$ from the coherence peak positions. Similar to $T_{BKT}$, the critical current decreases approximately linearly with decreasing twist angle, but at a larger rate, dropping to nearly zero value by 3.8\degree, despite $T_{BKT}$ remaining of order 100~mK.

The continuous evolution of the superconducting parameters suggests that the large and small twist angle superconducting pockets are smoothly connected and therefore probably represent the same ground state order. Across all devices the common observation is that the superconducting pocket remains adjacent to the AFM ordering.  At the same time, the superconducting pocket is observed whether or not there is a half band insulator, and whether or not the superconductor is coincident with the VHs.  Together, these findings indicate that superconductivity is spin-fluctuation mediated for all twist angles within the studied range. 

Fig.~2d shows an estimate of the superconducting correlation strength versus twist angle.  We plot the ratio of the experimental critical temperature $T_c$ to the Fermi temperature, $T_F$, and also the ratio of $T_{c}$ to the Drude weight $T_D$, where $T_{F}$ and $T_{D}$ are obtained from the calculated band structure
of the three-orbital Wannier model of t\wse (see Method Section). 
Both $T_c/T_F$ and $T_c/T_D$ show an upward trend with decreasing twist angle, reflecting an evolution to a more strongly correlated regime.  Nonetheless we note that $T_c/T_F$ and $T_c/T_D$ remain relatively small, reflecting values that suggest the superconductor remains in the weak to moderate correlation regime~\cite{uemuraCondensationExcitationPairing2004} (strongly correlated is typically taken as $T_{c}/T_{F}>1\%$).  Inset in Fig.~2d shows the  $T_{c}/T_{F}$ for our devices on an Uemura plot.  As twist angle decreases, the ratio shifts towards the strongly correlated regime (grey shaded area) but does not reach it.  The shift is consistent with expectations that in the low twist angle, low displacement field regime the band is flatter and therefore interactions are stronger/superfluid stiffness decreases.

\vspace{\baselineskip}
\noindent\textbf{Half-band insulator}

In Fig.~\ref{fig:fig3} we examine the evolution with twist angle of the resistivity at half filling  ($\nu=-1$). Figs.~3a-d compare the response between 3.8\degree and 4.8\degree devices. In both cases the device shows an excess resistance in the region associated with the Fermi surface reconstruction.  However, whereas the 4.8\degree device shows a metallic temperature dependence at all displacement field values (Fig.~3d), the 3.8\degree exhibits clear metal-to-insulator transitions (MIT) (Fig.~3b), with a sharp change in the sign of the temperature  dependence at around $T\sim8$~K when $0.1$~V/nm $\lesssim |D|\lesssim0.3$~V/nm. 

Fig.~3e-f shows the twist angle evolution of the temperature dependent resistivity across all measured devices at $\nu = 1$.  In Fig.~3d, the resistance versus temperature is compared, where each curve was acquired at half-band filling, and at a displacement field corresponding to the resistance maximum at 1.5 K.  The gap energy is deduced from thermal activation measurement (Fig.~3f) and plotted versus angle (open circles in Fig.~3g). Also plotted in Fig.~3g are calculated gap values (closed circles) obtained from Hartree-Fock simulations for three different interaction strengths, parametrized by dielectric constants $\varepsilon$. The angle dependence of the experimental and calculated gaps show good qualitative agreement. Quantitatively, the calculated gap values show a significant sensitivity to the choice of dielectric constant and screening length~\cite{munoz-segoviaTwistangleEvolutionIntervalleycoherent2025b} (see also Fig.~SI~9), but otherwise consistently increase with decreasing twist angle, in agreement with experiment.

In \moire systems with strong electron interactions, a Mott insulator is expected at half filling of the \moire mini-band. Normally the Mott insulator is characterized by a $d\rho/dT <0$ behaviour over a relatively large temperature range, similar to a trivial band insulator. Additionally, for twisted bilayer TMDs, prior theoretical work found that in the strong interaction regime the Mott insulator is expected to show little displacement field dependence extending from $D=0$ and persisting through the layer polarization regime~\cite{ryeeSitepolarizedMottPhases2025}. Both expectations are in stark contrast to our experimental observations where the half filling gap appears only in a narrow range of displacement field, corresponding to where the VHs approaches half filling, and with the insulating behaviour onsetting sharply below a critical temperature.  This behaviour suggests that even at the smallest angles measured, the system is in an intermediate regime of moderate interactions.

\vspace{\baselineskip}
\noindent\noindent\textbf{Phase diagram versus twist angle}

Fig.~\ref{fig:fig4}~a-d summarizes the phase diagram evolution across the four devices highlighted in Fig.~1.  We identify the superconducting (blue), insulator (brown), and two antiferromagnetic metallic (M1 and M2, dark and light yellow) phase boundaries, based on our low temperature transport measurements. The two metallic antiferromagnetic phases are distinguished by differences in the longitudinal and Hall magnetoresistance (see Fig.~SI~5).

In Fig.~4e-h we show a theoretically calculated phase diagram for each twist angle. The calculations use the functional renormalization group (FRG) to obtain the leading instabilities of a theoretical model of tWSe\textsubscript{2} based on dual-gated Coulomb interactions and a realistic three-orbital tight-binding model that captures the twist angle and doping dependence of the intricate quantum geometry and fermiology of the relevant bands~\cite{fischerTheoryIntervalleycoherentAFM2024}. 
We find superconducting (SC, blue) and magnetic density wave (DW, orange-red) instabilities. The magnetic orderings are further distinguished by color-coding the leading momentum transfer of the magnetic susceptibility $\chi^{\text{DW}}(\bvec q)$. The theoretical phase diagram is determined from the leading divergence and the instability temperature is inferred from the RG scale $\Lambda_{\mathrm{c}}$ at which the interaction reaches a critical value proportional to the moiré bandwidth at the respective twist angle, see Methods for details.

The FRG calculations reproduce pivotal aspects of the experimental phase diagram and yield microscopic insight into the evolution of magnetic and SC order as function of twist angle. At large twist angles (5\degree and 4.8\degree), the FRG predicts inter-valley coherent AFM order ($\bvec q=K$, IVC-AFM) and SC order. The order occurs for densities and displacement fields at which the Fermi level is close to the Van Hove line. The maximal transition temperature of the superconductor occurs at the point at which superconductivity gives way to IVC-AFM order. The point of the maximal SC transition temperature is also the point of maximal density of states within the superconducting pocket. Following the Van Hove line to lower displacement fields and densities closer to $\nu=-1$, we find that the density of states decreases and the nesting becomes less good, so that instabilities are not predicted at any reasonable RG scale.

In this twist angle regime, IVC-AFM order occurs at incommensurate densities $\nu<-1$ where the commensurate order cannot fully gap the Fermi surface. The resulting partially gapped Fermi surface~\cite{munoz-segoviaTwistangleEvolutionIntervalleycoherent2025b,fischerTheoryIntervalleycoherentAFM2024} explains the metallic resistive behavior~\cite{guoSuperconductivity50degTwisted2025,ghiottoStonerInstabilitiesIsing2024}. The `fish-tail' excess longitudinal resistance that refers to the two resistivity maxima emerging along a constant displacement field cut within the IVC-AFM region as visible in Fig.~\ref{fig:fig1}~d, can thus be understood from the the interplay of IVC-AFM order and Fermi surface geometry, leading to an intricate structure of partial gap opening and splitting of the VHs.

In the smaller twist angle samples, AFM and SC order move to smaller densities and displacement fields exhibiting an increasing asymmetry with respect to the Van Hove line~\cite{fischerTheoryIntervalleycoherentAFM2024}.
Driven by stronger correlations relative to the moiré bandwidth, IVC-AFM order occurs over increasingly larger domains in the phase diagram. Additionally, when the IVC-AFM order overlaps $\nu=-1$, the Fermi surface becomes fully-gapped in the ordered phase as confirmed by self-consistent Hartree-Fock calculations, see brown points in Fig.~\ref{fig:fig4}~g-h. At low twist angles, the displacement of the SC phase from the Van Hove line increases. Comparing the FRG simulations and the shape of IVC-AFM ordered regimes (grey dashed line) with the experimental phases M1 (M2), we can further identify the latter as hole-doped (electron-doped) IVC-AFM states featuring reconstructed metallic behavior and surrounding the IVC-AFM insulator at $\nu=-1$~\cite{xiaSimulatingHightemperatureSuperconductivity2025}. The domains of IVC-AFM order do not extend to zero displacement field such that SC order can prevail below.
The SC states for $\theta=4.2^{\circ}$ and $\theta=3.8^{\circ}$ are shrunk to small pockets in the vicinity of $\nu=-1$ and appear adjacent to the IVC-AFM insulator, reminiscent of the superconducting phase of rhombohedral graphene systems proximitized by TMD layers~\cite{pattersonSuperconductivitySpinCanting2025,yangImpactSpinOrbitCoupling2025}.
The decreased size of the SC pockets and decreased transition temperature arises in the calculation from the decrease of the DOS away from the VH line.

\vspace{\baselineskip}
\noindent\noindent\textbf{Discussion and conclusions}

Our study maps the evolution of the electronic properties of \twse as a function of twist angle, carrier density, displacement field and temperature and compares the experimental findings to theory.  All devices exhibit broad regions of apparently conventional metallic behavior,  with  antiferromagnetism and superconductivity at low temperatures in particular regimes of displacement field and carrier density.  Superconductivity always occurs adjacent to the magnetic phase, with the highest transition temperature at the boundary of the magnetic phase.  This remains true independent of the VHs position in the band or whether the Fermi surface is gapped at half filling.
The smooth evolution of properties with twist angle, and close association of the superconductivity to the magnetic phase at all twist angles,  strongly suggests that the observed superconductivity is mediated by spin fluctuations over the full angle range studied.

At large twist angles, antiferromagnetism and superconductivity are observed near regions where the Fermi surface intersects a VHs in the density of states. FRG calculations find both the magnetic density wave and superconducting states in approximately the same, experimentally-observed, locations in the phase diagram. In both experiment and theory we find that the Van Hove-driven antiferromagnetism incompletely gaps the Fermi surface. The magnitude of the resistivity is small. This phenomenology is consistent with a weak coupling picture in which the antiferromagnetism is driven by enhanced susceptibility (Stoner picture) and the superconductivity is mediated by spin fluctuations.

As the twist angle decreases, the AFM ordering shifts to lower density and displacement field, and it becomes a fully gapped insulator at $\nu = 1$. These features are also found in the calculations, and are understood theoretically as a combination of the evolution of the Fermiology with twist angle, a relative increase in the correlation strength, and a commensurability effect as the region of antiferromagnetism overlaps $\nu=-1$ more strongly. Although the evolution with decreasing twist angle, in both experiment and theory, suggests an evolution from weak coupling towards strongly coupled Mott physics, evidence suggest that even at the smallest twist angle studied, 3.8\degree, the sample remains in the crossover intermediate-coupling region.   

The twist-angle evolution of the superconductivity provides interesting insights into its origin. At large twist angles, superconductivity is robust, with a reasonably sharply defined transition and a resistance that goes to zero at base temperature. As noted, the phenomenology is consistent with a weak-coupling BCS picture in which the pairing glue is provided by spin fluctuations. However both the transition temperature and region of phase space where superconductivity is observed shrink with twist angle. Experimentally, at small twist angles, where insulating phases are stabilized, superconductivity becomes fragile and often manifests as a marginal or failed SC state (see Figs.~SI~4,5), suggesting a first order SC-insulating AF phase competition. An important parameter characterizing  the superconducting state is the superfluid stiffness, $\rho_s$. The ratio $T_c/\rho_s$ determines the position of the material on the `Uemura plot' of exotic superconductors.  The band theory values of $\rho_s$ range from 350~K ($5^\circ$) to 150~K ($3.65^\circ$), in all cases almost three orders of magnitude greater than the transition temperatures, placing the materials, at all twist angles, well on the `BCS' side of the BCS-BEC crossover. Of course, interactions will renormalize the superfluid stiffness, but renormalizations by factors of more than 10 do not occur without extreme fine tuning. We therefore believe that the superconductivity is in the BCS limit for all twist angles.

\section{Acknowledgments}
\begin{acknowledgments}
The research on superconductivity in tWSe2 structures was primarily supported as part of Programmable Quantum Materials, an Energy Frontier Research Center funded by the US Department of Energy, Office of Science, Basic Energy Sciences, under award no. DE-SC0019443.  WSe\textsubscript{2} was synthesized by J.H., K.B. and L.H. under the support of the Columbia University Materials Science and Engineering Research Center (MRSEC), through NSF grants DMR-2011738. K.W. and T.T. acknowledge support from the JSPS KAKENHI (Grant Numbers 21H05233 and 23H02052), the CREST (JPMJCR24A5), JST and World Premier International Research Center Initiative (WPI), MEXT, Japan. A.F. and D.M.K. acknowledge funding by the DFG  within the Priority Program SPP 2244 ``2DMP'' -- 443274199. J.H. and C.R.D. acknowledge additional support from the Gordon and Betty Moore Foundation’s EPiQS Initiative, Grant GBMF10277. The Flatiron Institute is a division of the Simons foundation. 
\end{acknowledgments}

\section{Author Contributions}
Y.G. and J.C. fabricated the devices. Y.G., J.C., J.P., and C.R.D. performed the electronic transport measurements and analyzed the data.  A.F., D.M., and L.K. performed theoretical modeling under the supervision of D.M.K. and A.J.M. L.H. grew the \wse crystals under the supervision of J.H. and K.B.; K.W. and T.T. grew the hexagonal boron nitride crystals. Y.G., A.J.M. and C.R.D. wrote the manuscript with input from all authors. C.R.D. supervised the project.

\section*{Competing financial interests}
The authors declare no competing financial interests.

\bibliography{angle_evolution}

\section{Methods}

\subsection{Device fabrication}
Device images and structure illustrations are shown in Fig.~SI~1. We use similar fabrication process described the previous literature~\cite{guoSuperconductivity50degTwisted2025,packChargetransferContactsMeasurement2024,xuObservationIntegerFractional2023c}. The device consists of \twse with graphite as top and bottom gates, hBN as the dielectric spacer. Devices at 5\degree, 4.8\degree, 4.3\degree, 4.1\degree and 3.8\degree are made with graphite contacts, and \rucl flake is used to heavily dope the contacts~\cite{packChargetransferContactsMeasurement2024}. Device at 4.2\degree is made with TaSe$_2$\cite{xuObservationIntegerFractional2023c} as contacts to \twse. The two dimensional material flakes, including hBN, graphite, \wse are exfoliated on SiO$_2$/Si chips, \rucl flakes are exfoliated on chips pre-coated with hexamethyldisilane (HMDS) to make it hydrophobic, and TaSe$_2$ flakes are exfoliated inside a glovebox. The flakes are picked in sequence with thin polycarbonate film~\cite{wangOneDimensionalElectricalContact2013}. The \wse flake is pre-cut into two pieces with an atomic force microscopy tip. For fabrication on the stack, metal contact gates are first deposited for \rucl doped contact devices. Then metal contacts are deposited onto the top gate, bottom gate, graphite contacts with the same etch window used to etch through the hBN flake down to the graphite layer. Finally, we etch apart the contacts and shape the stack into a half Hall bar geometry with contacts on one side.

\subsection{Measurements}
Transport measurements were performed in a variable temperature cryostat with a base temperature of 1.6 K and a dilution fridge. Four and two terminal resistance measurements were carried out using a low-frequency lock-in technique at frequencies ranging from 6 Hz-25 Hz. Characterization for  superconducting region is done with a current source bias of 1 nA - 5 nA. $dV/dI$ measurements were done using lockin SR860 to provide an AC signal on top of the DC signal. The current signal is amplified using a SR570 and measured with SR860s and a nanovoltmeter. 

\subsection{Identification of max $T_c$}
In order to identify the position of the max $T_c$ point in the $n-D$ phase diagram, we go to a temperature near $T_c$ where all resistance inside the superconducting pocket becomes non-zero. We did multiple fine line sweeps across the superconducting pocket (the shape of the pocket was previously identified at base dilution fridge temperature) and found out the density point which shows the lowest resistance. 

\subsection{Multi-orbital Wannier model}

Starting from a continuum model of \wse{}~\cite{devakul2021magic}, the bandstructure and quantum geometry of the moiré flat bands are captured by constructing a multi-orbital Wannier basis~\cite{crepelBridgingSmallLarge2024a,fischerTheoryIntervalleycoherentAFM2024} comprising one triangular lattice site $T$ in the MM stacking regions of the triangular superlattice (Wyckoff position 1a) as well as two honeycomb lattice sites $H_{1,2}$ in the XM/MX stacking regions (Wyckoff positions 2a,b). 
The long-ranged Coulomb interaction specific to the dual-gated device architecture 
\begin{equation}
    V(r) = 4V_0\sum_{k=0}^\infty K_0\left[ (2k+1)\pi\,\frac{\sqrt{r^2+a^2}}{\xi} \right] \,.
\label{eq:dual-gated-coulomb}
\end{equation}
is then projected onto the Wannier basis following the procedure outlined in Ref.~\cite{fischerTheoryIntervalleycoherentAFM2024}. $K_0$ denotes a modified Bessel function of the second kind, $\xi = 100$ \AA{} is the gate distance and $\alpha = 14.40$\,eV \AA{} is the fine-structure constant. The interaction parameters $a = \alpha/(\epsilon U)$ and $V_0=\alpha/(\epsilon\xi)$ are determined by the dielectric constant $\epsilon=14$ and the on-site interaction strength $U = 3$ eV.
This results in a multi-orbital moiré Hubbard Hamiltonian
\begin{equation}
\begin{split}
H=\sum_{\nu,\bvec R,\bvec R'}\sum_{X,X'} t^{\nu}_{\bvec R X,\bvec R'X'}c^{\nu\,\dagger}_{\bvec R X} c^{\nu\vdagger}_{\bvec R' X'}  \\
+ \sum_{\nu,\bvec R}\,\sum_{X\in\{T, H_1,H_2\}} \frac{U_X}2 n_{\bvec RX}^\nu n_{\bvec RX}^{\bar\nu} \,.
\end{split}
\label{eq:fullham}
\end{equation}
Here, $\nu$ denotes the locked spin-valley degree of freedom, $X,X' \in \{T, H_1, H_2 \}$ orbital, and $\bvec R, \bvec R'$ Bravais lattice vectors. The operator $c_{\bvec RX}^{\nu\,(\dagger)}$ destroys (creates) an electron with spin/valley $\nu$ in the orbital $X$ on site $\bvec R$, and $n^\nu_{\bvec RX} = c^{\nu\,\dagger}_{\bvec RX} c^{\nu\vdagger}_{\bvec RX}$ is the density operator. For a detailed overview of the interaction values $U_X$ and calculations of non-interacting properties including the density of states (DOS), Fermi temperature $T_F$ and bandwidth, see Supplementary Material.

\subsection{Functional renormalization group calculations}

We address electronic order from first-principles in the weak-to-moderate interacting regime by employing the functional renormalization group (FRG) that represents a well-established method to predict particle-particle and particle-hole instabilities in an unbiased manner~\cite{metznerFunctionalRenormalizationGroup2012}.
The FRG smoothly connects the non-interacting model to an effective low-energy theory near the Fermi level by successively integrating out high-energy degrees of freedom. This is achieved by introducing a flowing energy cutoff (the renormalization group scale) $\Lambda = \infty \to 0$, above which the effect of quantum fluctuations is systematically included in the effective vertex functions $\Gamma^{(2n)}$. By perturbatively expanding possible scattering processes, the FRG allows for an unbiased treatment of electronic order and symmetry-breaking transitions in the pairing, charge and magnetic channel. This makes FRG the distinguished method to study the competition of correlated states in the crossover regime from weak-to-moderate interactions in \twse.

To track the entire orbital/band and (incommensurate) momentum dependence of the multi-orbital Wannier model, we employ the static four-point truncated unity FRG (TUFRG) approximation~\cite{profeDivERGeImplementsVarious2024}. 
The level-2 truncated FRG flow equations for the two-particle vertex function $\Gamma = \Gamma^{(4)}$ read
\begin{equation}
\partial_{\Lambda} \Gamma^{\Lambda} = \sum_\gamma \partial_{\Lambda} \gamma^{\Lambda} \,, \quad \partial_{\Lambda} \gamma^{\Lambda} =  \Gamma^{\Lambda} \circ \partial_{\Lambda} L^{\gamma, \Lambda} \circ \Gamma^{\Lambda} \,,
\label{eq-frg}
\end{equation}
where $\gamma \in \{P,D,C\}$ denotes the decomposition into two-particle reducible interaction channels that capture fluctuations accumulated in the particle-particle ($P$), direct particle-hole ($D$) and crossed particle-hole ($C$) channel. This allows to treat all couplings on equal footing and (at one loop order) accounts for mutual feedback between the different interaction channels that are renormalized during the flow.
In static four-point (TU)FRG, the onset of an ordered phase is signaled by the divergence of the two-particle vertex $\Gamma^{(4)}$ at a particular (critical) scale $\Lambda_\mathrm{c}$. The leading fermionic bilinear in the divergent channel is obtained by an eigenvalue decomposition of the channel-specific vertex $\gamma$
\begin{equation}
\gamma_{\kappa \kappa'}(\bvec q_\gamma) = \sum_i \phi^L_{\kappa,i}(\bvec q_\gamma) \, \gamma_i(\bvec q_\gamma) \, \phi^R_{\kappa',i}(\bvec q_\gamma) \,,
\label{eq-fermionic-bilinear}
\end{equation}
where $\kappa = (\bvec k_\gamma, X, \nu_1,\nu_2)$ is a multi-index comprising the fermionic momentum variable $\bvec k_\gamma$, Wannier function index $X$, and spin/valley $\nu_{1,2}$. The left/right eigenvectors to the channels' eigenvalues $\gamma_i$ are denoted by $\phi^{L/R}_{\kappa,i}$. They provide insight into the momentum, spin/valley and orbital structure of the order parameter. Since we choose the sharp frequency cutoff as regulator (as implemented in the divERGe library~\cite{profeDivERGeImplementsVarious2024}), the critical scale $\Lambda_c$ serves as a proxy for the critical temperature of the transition. 

\subsection{Hartree-Fock calculations}

We perform self-consistent zero-temperature Hartree-Fock (HF) calculations in a $\sqrt{3}\times\sqrt{3}$ supercell for the multi-orbital Hubbard model of Eq.~\eqref{eq:fullham} (See Ref.~\cite{munoz-segoviaTwistangleEvolutionIntervalleycoherent2025b} for additional details on the HF calculations). The $\sqrt{3}\times\sqrt{3}$ supercell captures the dominant $\boldsymbol{q}=K$ intervalley and $\boldsymbol{q}=0$ intravalley instabilities predicted by the weak-coupling analysis~\cite{munoz-segoviaTwistangleEvolutionIntervalleycoherent2025b} and confirmed by the FRG calculations~\cite{fischerTheoryIntervalleycoherentAFM2024}. In order to directly compare the HF to the FRG calculations, which renormalizes the vertex but not the self-energy, we set to zero the symmetric Hartree renormalization of the onsite energies of the three orbitals, which would otherwise change the parent band structure. To effectively reproduce the emergent screening of the Coulomb interaction by particle-particle scattering as captured in FRG, we use a dielectric constant $\epsilon_{\mathrm{HF}} = 2.2 \epsilon_{\mathrm{FRG}}$ in quantitative agreement with Ref.~\cite{fischerTheoryIntervalleycoherentAFM2024}.

\section{Data availability}
The data that support the plots within this paper and other findings of this study are available from the corresponding author upon reasonable request.

\end{document}